\documentstyle[aps,draft,epsf,amssymb,amscd]{revtex}

 \newcommand{\eins}{\mbox{$1 \hspace{-1.0mm}  {\bf l}$}}
 
 \newcommand{\be}{\begin{equation}}
 \newcommand{\ee}{\end{equation}}
 \newcommand{\bea}{\begin{eqnarray}}
 \newcommand{\eea}{\end{eqnarray}}
 \newcommand{\half}{\mbox{$\textstyle \frac{1}{2}$}}
 \newcommand{\shalf}{\mbox{$\textstyle \sqrt{\frac{1}{2}}$}}
 \newcommand{\seight}{\mbox{$\textstyle \sqrt{\frac{1}{8}}$}}
 \newcommand{\ket}[1]{ | \, #1  \rangle}
 \newcommand{\bra}[1]{ \langle #1 \,  |}
 \newcommand{\proj}[1]{\ket{#1}\bra{#1}}

 \newcommand{\ok}{|0\rangle}
 \newcommand{\uk}{|1\rangle}

 \newcommand{\Cdue}{\mathbb{C}^{\mbox{\scriptsize 2}}}

\begin{document}
                 \title{  Phase covariant quantum cloning  }
                 \author{Dagmar~Bru\ss $^1$,  Mirko Cinchetti $^2$,
                 G. Mauro D'Ariano $^2$ and Chiara~Macchiavello$^{2}$}
                 \address{
                 $^1$Inst. f\"{u}r Theoret. Physik, Universit\"{a}t Hannover,
                 Appelstr. 2, D-30167 Hannover, Germany\\
                 $^2$Dipartimento di Fisica ``A. Volta'' and INFM-Unit\`a 
                 di Pavia,
                 Via Bassi 6, 27100 Pavia, Italy}
                 \date{Received \today}
                 \maketitle
                  \begin{abstract}

We consider an $N\rightarrow M$ quantum cloning transformation acting 
 on pure two-level
states lying on the equator of the Bloch sphere.
An upper bound for its fidelity is presented, 
by establishing a connection between optimal phase covariant cloning  and
 phase estimation.
We  give the explicit form of a cloning transformation that achieves 
the bound for the case N=1, M=2, and find a link between this 
case and optimal eavesdropping in the quantum cryptographic scheme BB84.   

                 \end{abstract}
                 \widetext

\section{Introduction}

Perfect quantum cloning of a set of input states that contains at 
least two non-orthogonal states is impossible
\cite{wootters}.
However, it is interesting to study how well we can approximate a perfect 
cloning procedure.
We can expect different results depending on the set of input 
states considered. In particular, we expect that the smaller the set
of inputs,
i.e. the more information about the input is given,    
the better one can clone each of its states. 

We analyze the case of pure qubits, i.e. vectors of a 
two-dimensional Hilbert space $\mathcal{H}\simeq \Cdue$.
Optimal $N\to M$ cloning transformations 
(i.e. transformations which act on $N$ identical inputs and create $M$ outputs)
for the largest set of
input qubits, namely for qubits belonging to the whole Hilbert space,
have been recently proposed \cite{gm,bc,Wer}.
Since a crucial requirement for such transformations is that their efficiency
is the same for all input states, they were called universal cloning 
transformations.

In this paper we will analyse  cloning transformations that are optimal for
a restricted set of input states, namely pure states of the form
\begin{equation}
|\psi_{\phi}\rangle=\frac{1}{\sqrt{2}}\left[\ok+e^{i\phi}\uk\right]\;,
\end{equation}
where $\phi\in[0,2\pi)$ and $\{\ket{0},\ket{1}\}$ represent a basis for
a qubit. 
We call the qubits of this form ``equatorial'' because 
the $z$-component of their Bloch vector 
is zero, i.e. the Bloch vector is restricted to the 
intersection of the $xy$ plane with the Bloch sphere.
The parameter $\phi$ is the angle between the Bloch vector 
and the $x$-axis.
\par
Studying the restriction of the input set to the equator is motivated by 
physical implementations of quantum communication ideas
(all existing quantum cryptographic experiments
are using states that are on the equator, rather than states that span the
whole Bloch sphere) as well as by fundamental  
questions in quantum information processing.
As we will show in this paper, restricting to equatorial states makes the
cloning problem related to phase estimation. This connection can be exploited 
in order to derive bounds for the optimal cloning fidelity. 
As expected, restriction of the cloning symmetry improves the cloning
performance.

The paper is organised as follows. In Sect. II we describe the 
general operation of a 
phase covariant cloning transformation. In Sect. III we establish the 
connection 
between phase covariant cloning and phase estimation, and prove an upper
bound on the fidelity of an $N\to M$ phase covariant cloner acting on
equatorial qubits. In Sect. IV we derive the explicit form of the $1\to 2$
 cloning transformation for equatorial qubits that saturates the bound,
and point out a connection to eavesdropping in quantum cryptography.

\section{Phase covariant cloning transformations}  

 In this section we consider cloning transformations with the 
requirement that the fidelity is the same for any equatorial qubit, 
i.e. it does not depend on the value of the phase $\phi$.
We call such cloners ``phase covariant cloners'' (pcc).

We describe the action of an $N\to M$ phase covariant cloner on the $N$
input 
qubits by means of a completely positive (CP) map $T_{NM}$ \cite{Kra}.
We will consider only pure input states of the form
$\proj{\psi_\phi}^{\otimes N}$, namely product
states made of $N$ identical copies. 
The output of the map
is  generally a mixed state $\rho_M$ of the $M$ output qubits.
In order to guarantee that all the output copies are described by the same
density operator we require that $\rho_M$ is supported on the symmetric 
subspace of the total Hilbert space of the $M$ output qubits
(the symmetric subspace is defined as the space spanned by all pure states  which 
  are invariant under any permutation of the constituent qubits).
The density operator describing the state of each output qubit is given by
\begin{equation}
\rho^{out}=R[T_{NM}(\ket{\psi_\phi}\bra{\psi_\phi}^{\otimes N})]\;,
\label{out}
\end{equation}
where $R$ denotes the partial trace over all but one output qubits.
The phase covariance condition corresponds to imposing the following
requirement on the operation of the cloning map 
\begin{equation}
U_{\chi}\rho^{out}U^\dagger_{\chi}=
R[T_{NM}(U_{\chi}^{\otimes N}
\ket{\psi}\bra{\psi}^{\otimes N}
{U_{\chi}^{\dagger \otimes N}})]\;
\label{covariance}
\end{equation}
for any pure state $\ket{\psi}$ and all unitary phase shift operators 
$U_{\chi}=\exp{\left[-\frac{i}{2}
(\sigma_z-\eins)\chi\right]}$, where $\chi\in[0,2\pi)$ and $\sigma_z$ 
is the  Pauli operator $diag\{ 1,-1 \}$.

We define the quality of the cloning transformation in terms of the fidelity 
between the reduced density operator of each output copy and the input state
$\ket{\psi_\phi}$
\begin{equation}
F=\bra{\psi_\phi}\rho^{out}\ket{\psi_\phi}\;.
\end{equation}
In the appendix we show that without loss of generality 
  any phase covariant cloning transformation 
can be completely described in terms of two shrinking factors $\eta_{xy}(N,M)$
and $\eta_z(N,M)$. The former 
describes the shrinking of the component of the Bloch
vector lying in the $xy$ plane of the Bloch sphere, the latter the
shrinking of the component along the $z$ direction, namely the state
of each output copy is
\begin{eqnarray}
\rho^{out}=\frac{1}{2}[\eins+\eta_{xy}(N,M)(s_{x}\sigma_x
+s_{y}\sigma_y)+\eta_z(N,M)s_{z}\sigma_z]\;,
\end{eqnarray}
where $s_i$ are the components of the Bloch vector of the initial state
$\ket{\psi}$ of each of the $N$ input copies. Therefore, 
for equatorial qubits, the cloner
leads to an isotropic shrinking, namely the density 
operator of each output copy (\ref{out}) is given by
\begin{equation}
\rho^{out}=\eta_{xy}(N,M)\ket{\psi_{\phi}}\bra{\psi_{\phi}}
+\frac{1}{2}[1-\eta_{xy}(N,M)] \eins\;,
\end{equation}
where $\eins$ is the identity operator.
Thus, for equatorial qubits the action of a phase covariant cloner is
completely specified in terms of the equatorial shrinking factor
$\eta_{xy}(N,M)$ and the fidelity is $F_{pcc}(N,M)=(1+\eta_{xy}(N,M))/2$.

\section{Optimal cloning of equatorial qubits}

In this section we derive an upper bound for the shrinking factor 
$\eta_{xy}(N,M)$ of a phase covariant cloner for equatorial qubits.
Our derivation is similar to the one of universal cloners \cite{bc}.
It is based on the concatenation property of phase covariant cloners 
and on the link to phase estimation, as shown in the following.

\subsection{Concatenation of phase covariant cloners}

We concatenate two phase covariant cloners as follows. 
The first is an $N\to M$ cloner acting on $N$ equatorial qubits,
the second one acts on the output state $\rho_M$ of the $M$ output
qubits of the first cloner and gives $L$ output copies. 
We show in the following that the sequence
of these two cloning transformations is a phase covariant cloner with
a shrinking factor $\eta_{xy}$ for the $xy$ plane  that is the multiplication 
of the shrinking factors $\eta_{xy}$ of the two separate cloners, namely
\be
\eta_{xy}(N,L)=\eta_{xy}(N,M)\cdot\eta_{xy}(M,L)\ \ .
\ee

In order to prove the above property we exploit the  decomposition
of a density operator supported on the symmetric subspace \cite{Wer},
\be
\rho_M=\sum_i\beta_i\proj{\psi_i}^{\otimes M}\ , 
\label{decomp}
\ee
with 
$\beta_i \in \mathbb{R}$ (not necessarily positive) and
$\sum_i\beta_i=1$.

Using the shrinking character of the phase covariant cloning transformation
described in the previous section and  the linearity of the 
cloning map we can write the following conditions for the output of the
$N\to M$ cloner acting on $N$ pure qubits in the generic pure state 
$\ket{\psi}$ with (unit-length) Bloch vector $\vec{s}$:

\begin{eqnarray}
 & & \sum_{i}\beta_{i}s_{xi}=\eta_{xy}(N,M) s_x\ \ ,\nonumber\\
 & & \sum_{i}\beta_{i}s_{yi}=\eta_{xy}(N,M) s_y\ \ ,\nonumber\\
 & & \sum_{i}\beta_{i}s_{zi}=\eta_z(N,M) s_z\;,
\label{condizio}
\end{eqnarray}
where $s_{xi}$ denotes the $x$-component of the
Bloch vector  of state $\proj{\psi_i}$, and accordingly for 
$y,z$.

The reduced density operator describing each of the $L$ copies at the output 
of the second cloner is given by

\begin{eqnarray}
R[T_{ML}(\rho_M)]=\sum_i\beta_i R[T_{ML}(\proj{\psi_i}^{\otimes M})]=
\sum_i\beta_i\left\{\frac{1}{2}[\eins+\eta_{xy}(M,L)(s_{xi}\sigma_x
+s_{yi}\sigma_y)+\eta_z(M,L)s_{zi}\sigma_z]\right\}\;.
\end{eqnarray}

By using eqs. (\ref{condizio}) the above expression takes the form

\begin{eqnarray}
R[T_{ML}(T_{NM}(\proj{\psi}^{\otimes N}))]=
\frac{1}{2}[\eins+\eta_{xy}(N,M)\cdot\eta_{xy}(M,L)(s_{x}\sigma_x
+s_{y}\sigma_y)+\eta_z(N,M)\cdot\eta_z(M,L)s_{z}\sigma_z]
\end{eqnarray}
namely the concatenation property holds.
For input  qubits from the equator the Bloch vector of each copy at the output of
the two cloners is simply shrunk in the $xy$ plane by the factor 
$\eta_{xy}(N,M)\cdot\eta_{xy}(M,L)$.

\subsection{Phase covariant cloning and phase estimation}

We will now prove the following connection between phase covariant cloners and
phase estimation of equatorial qubits:
\be
\eta_{xy}^{opt}(N,\infty)=\overline{\eta}^{opt}_{pe}(N)\;.
\label{equal}
\ee
The quantity $\eta_{xy}^{opt}(N,M)$ is the shrinking factor 
in the $xy$ plane of the optimal 
$N\rightarrow M$ phase covariant cloner, while 
$\overline{\eta}_{pe}^{opt}(N)$ is 
the shrinking factor of the reconstructed reduced density operator after 
performing phase  estimation (pe) on $N$ equatorial qubits.
\par
The aim of phase estimation is to find the optimal strategy to estimate the
value of the phase $\phi$. This is described in terms of a positive-operator
valued measure (POVM), namely  $d\mu(\phi_*)$, where $\phi_*$
is the estimated value of the phase, $d\mu(\phi_*)\geq 0$ 
and $\int\frac{d\phi_*}{2\pi}d\mu(\phi_*)=\eins$.
The outcome of each instance of measurement provides, with probability 
$p(\phi|\phi_{*})=Tr\left[d\mu(\phi_{*})|\psi_{\phi}\rangle
\langle\psi_{\phi}|\right]$, the ``candidate'' $|\psi_{\phi *}\rangle$ for  
$|\psi_{\phi}\rangle$. The fidelity of phase estimation can be calculated
from the outcomes of the measurement as
\begin{eqnarray}
\overline{F}_{pe}(N) & = & \int\frac{d\phi_{*}}{2\pi}p(\phi|\phi_{*})
|\langle\psi_{\phi}|\psi_{\phi_{*}}\rangle|^{2}=\nonumber\\
& = & \langle\psi_{\phi}|\overline{\varrho}_{\phi}|\psi_{\phi}\rangle\ \ ,
\label{nondipdaphi}
\end{eqnarray}
where $\overline{\varrho}_{\phi}=\int\frac{d\phi_{*}}{2\pi}p(\phi|\phi_{*})
|\psi_{\phi_{*}}\rangle\langle\psi_{\phi_{*}}|$ is the reconstructed density
operator.
For covariant phase estimation the fidelity 
does not depend on ${\phi}$, thus for the optimal procedure 
$\overline{\varrho}_{\phi}$ can also be written as
\begin{equation}
\overline{\varrho}_{\phi}=\overline{\eta}_{pe}(N)|\psi_{\phi}
\rangle\langle\psi_{\phi}|+\frac{1}{2}\left[1-\overline{\eta}_{pe}(N)
\right]\eins\;,
\label{rhos}
\end{equation}
namely  the input state is shrunk by the factor
$\bar\eta_{pe}(N)=2\bar F_{pe}(N)-1$.   

The fidelity for optimal covariant phase estimation of equatorial qubits, 
 derived in \cite{dbe}, takes the form
\begin{equation}
\overline{F}^{opt}_{pe}(N)=\frac{1}{2}
+\frac{1}{2^{N+1}}\sum_{l=0}^{N-1}\sqrt{{N \choose l}{N \choose l+1}}\;.
\label{Fmediaottima}
\end{equation}  
In order to prove Eq. (\ref{equal}) we first notice that after performing 
optimal
phase estimation on $N$ equatorial qubits all in state $\ket{\psi_\phi}$ we 
can prepare a state of $L$ qubits, supported on the symmetric subspace,
where each qubit is described by the reduced density operator
(\ref{rhos}). This procedure can be viewed as a phase covariant cloner and
therefore it cannot perform better than the optimal $N\to L$ phase covariant 
cloning transformation. Thus we can write the inequality
\be
\overline{\eta}^{opt}_{pe}(N) \leq \eta_{xy}^{opt}(N,L)\;,
\label{in1}
\ee
which holds for any value of $L$, and in particular for $L\to\infty$.

We will now prove the opposite inequality (which holds for 
$L\to\infty$ only):
 we concatenate a phase-covariant $N\to L$ cloner,
acting on equatorial qubits,  with a 
subsequent optimal {\em state} estimation (se) procedure 
(note that state estimation 
on qubits includes also an estimate of their phase). 
The whole procedure can be 
seen as a {\em phase} estimation performed on the input 
$\proj{\psi_{\phi}}^{\otimes N}$, with fidelity 

\begin{displaymath}
\overline{F}_{pe}(N)=
\langle\psi_{\phi}|\Lambda_L(\rho_L)
|\psi_{\phi}\rangle \ ,\qquad
\Lambda_L(\rho_L)=
\sum_{\mu}{\mbox{Tr}}\left[P_{\mu}\rho_{L}
\right]|\psi_{\mu}\rangle\langle\psi_{\mu}| \ ,
\end{displaymath}
where  $\rho_L$ is the output of the cloner and 
$\Lambda_L(\rho_L)$ is the CP-map of the state estimation of $L$ qubits,
$\{ P_{\mu}\}$ represents 
the set of optimal POVM's for  state estimation of $L$ qubits 
\cite{Mas,dbe} and $\ket{\psi_\mu}$ denotes the candidate for
$\ket{\psi}$ when performing the 
measurement $P_\mu$.
Since $\rho_L$ is supported on the symmetric subspace we use again the
decomposition $\rho_L=\sum_{i}\beta_i\proj{\psi_i}^{\otimes L}$ and 
obtain
\begin{eqnarray}
\overline{F}_{pe}(N) & = & \sum_{i}\langle\psi_{\phi}|\beta_{i}
\Lambda_L(\ket{\psi_{i}}\bra{\psi_i})^{\otimes L}
|\psi_{\phi}\rangle=\nonumber\\ 
& = & \sum_{i}\langle\psi_{\phi}|\beta_{i}
\left[\overline{{\eta}}^{opt}_{se}(L)\ket{\psi_{i}}\bra{\psi_i}+
\frac{1}{2}\left(1-\overline{{\eta}}^{opt}_{se}(L)\right)
   \eins\right]
|\psi_{\phi}\rangle
\label{2}
\end{eqnarray}  
where 
the optimal shrinking factor for state estimation is given by
$\overline{{\eta}}^{opt}_{se}(L)=
(2\overline{{F}}_{se}^{opt}(L)-1)=\frac{L}{L+2}$ \cite{Mas}.
Taking the limit of (\ref{2}) for $L\rightarrow \infty$ we have
\begin{displaymath}
\begin{CD}
\overline{F}_{pe}(N)@>L\rightarrow\infty>>\sum_{i}\langle\psi_{\phi}|
\beta_{i}\ket{\psi_{i}}\bra{\psi_i}\psi_{\phi}\rangle=
\frac{1}{2}\left[\eta_{xy}(N,\infty)
+1\right]\ \ .
\end{CD}
\end{displaymath}
The concatenation of a phase covariant
cloner with a state estimation  cannot perform better than 
the optimal phase estimation, thus we can write
\begin{equation}
\eta_{xy}^{opt}(N,\infty)\leq\overline \eta^{opt}_{pe}(N).\label{3}
\end{equation}
The inequalities (\ref{in1}) and (\ref{3}) prove the equality (\ref{equal}).

\subsection{Bound for optimal phase covariant cloning}

\begin{figure}[hbt]
\setlength{\unitlength}{1pt}
\begin{picture}(500,300)
\epsfysize=10cm
\epsffile[72 230 540 560]{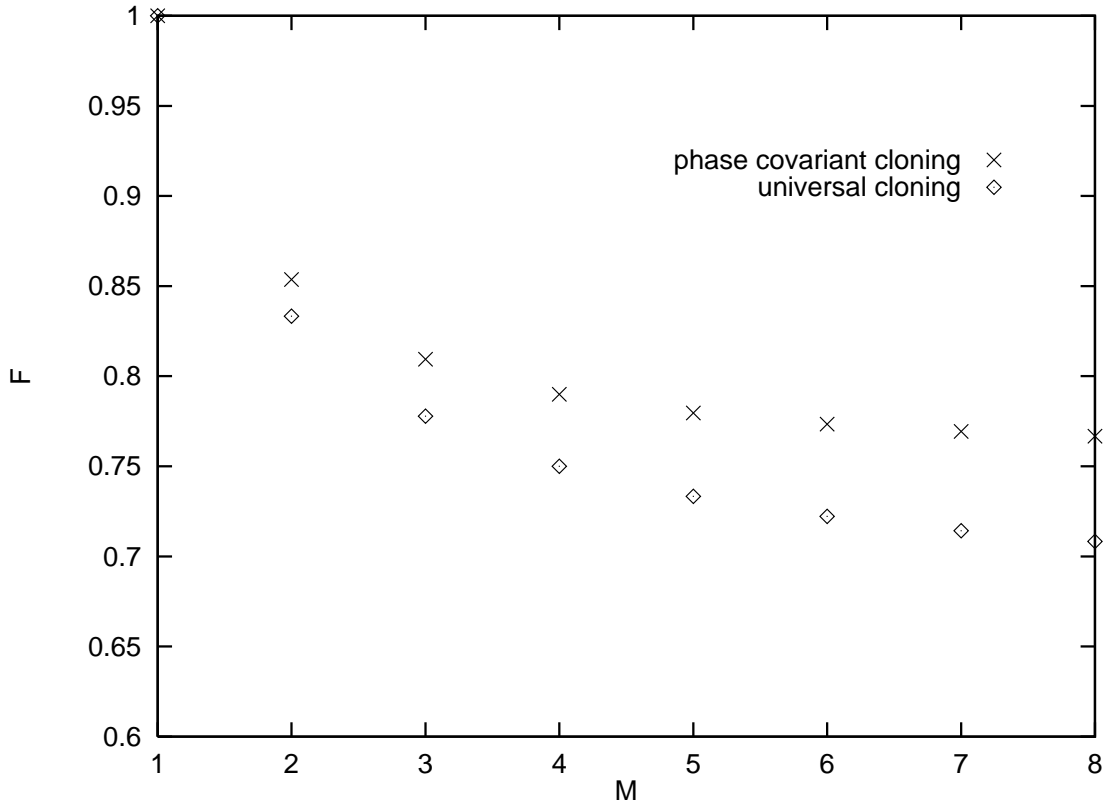}
\vspace{-0.2cm}
\end{picture}
\vspace{7cm}
\caption[]
        {\small Upper bound for the fidelity in phase covariant cloning
        compared with the optimal fidelity for universal cloning
        of qubits. Both sets of points
        are shown
        for a fixed number of inputs, $N=1$, as function of $M$, the number
        of outputs. For the limit $M \to \infty$ one finds from the
        formulae given in the text that $\tilde F_{pcc}(1,\infty)= 3/4$
        and $F^{opt}_{univ}(1,\infty)= 2/3$.
               }
\label{fideli}
\end{figure}

We now prove an upper bound for the fidelity of an $N\to M$ phase covariant
cloning transformation acting  on equatorial qubits.
We consider a phase-covariant cloner $T_{N\infty}$ 
that results from concatenating 
the two phase-covariant cloners $T_{NM}$ and  
$T_{{M\infty}}$. In this way we cannot obtain an $N\to \infty$ 
cloner that works better than the optimal one. Thus, by using the 
concatenation property of phase covariant cloners proven above we can write
\be
\eta_{xy}(N,M)\cdot\eta_{xy}(M,\infty)\le\eta_{xy}^{opt}(N,\infty)\;.
\label{nochnlabel}
\ee
In the sequence of the two cloners we  take the $M\to\infty$ as the 
optimal one in order to find the tightest
 upper bound for the
equatorial shrinking factor of a phase covariant $N\to M$
cloning transformation. We rewrite equation (\ref{nochnlabel}) as follows:
\begin{equation}
\eta_{xy}^{opt}(N,M)
\leq\frac{\eta_{xy}^{opt}(N,\infty)}{\eta_{xy}^{opt}
(M,\infty)}\;.
\label{boetaconc}
\end{equation}
By exploiting the connection to phase estimation in equation (\ref{equal}),
proven above, 
this bound takes
the form
\begin{eqnarray}
\eta_{xy}^{opt}(N,M)  \leq  \tilde{\eta}_{pcc}(N,M) &= &
\frac{\overline{\eta}^{opt}_{pe}(N)}
      {\overline{\eta}^{opt}_{pe}(M)}\nonumber\\
& = & 2^{(M-N)}\frac{\sum_{l=0}^{N-1}\sqrt{{N \choose l}{N \choose l+1}}}
{\sum_{j=0}^{M-1}\sqrt{{M \choose j}{M \choose j+1}}}\label{limsup}
\end{eqnarray}  
In Figure \ref{fideli} we show the upper bound
for the fidelity of phase covariant cloning and
the optimal fidelity
for a universal cloner. 
The two quantities are shown as a function of $M$
 for  fixed $N=1$. By varying N it is possible to see 
that
\be
\tilde\eta_{pcc}(N,M)>\eta^{opt}_{univ}(N,M)\quad \forall N<M \ \ ,
\ee
as expected.
Note that while in the case of universal cloning the explicit 
form of the CP map which achieves the bound is known \cite{gm},
 in the case of phase covariant cloners 
 acting on equatorial qubits we do not know
whether the bound (\ref{limsup}) can be achieved for 
general values of $N$ and $M$.
In the next section we present the cloning transformation which achieves 
the bound in the particular case N=1, M=2.

\vspace{-2cm}

\section{Optimal $1\rightarrow 2$ cloning of equatorial qubits}

In this section we present a constructive proof for 
 the best $1 \to 2$ cloning transformation acting on 
equatorial  qubits.
For convenience we  choose the  equator  in the $xz$ plane
instead of the $xy$ equator.
 (Note that optimality of the fidelity 
 must be independent from the choice of a 
particular basis.) Hence we consider equatorial states 
with real coefficients of the form
\be
\ket{\psi} = \alpha \ket{0} + \beta \ket{1} \ \ \ \ 
\text{with} \ \ \ \alpha, \beta \ \
\text{real}, \ \ \ \alpha^2+\beta^2=1 \ .
\label{state}
\ee
Our notation and method is inspired by \cite{oxibm}. We proceed as
follows:  first we derive the optimal cloner that takes only the four
BB84 states as input.  
Here we use the acronym BB84 for the quantum cryptographic protocol
described in \cite{bb84}. Remember that the four 
 BB84 states are given by:
                 \be
                 \ket{ 0}  ,\ \ 
                 \ket{ 1} ,\ \
                 \ket{\bar 0}  =   \shalf (\ket{0}+\ket{1})  ,\ \
                  \ket{\bar 1}   =    \shalf (\ket{0}-\ket{1})\ .
                  \ee
Then we will show that this transformation leads to
the same fidelity for {\em any } input from the equator.
Therefore we
have also found the best transformation that takes all states from
the equator as input. (If we could find a better one on the whole equator 
it would have to be better than the optimal one for the BB84 states.)
\par
We start from a general symmetric ansatz for the unitary transformation on 
the input qubit, blank qubit and ancilla, written in this order:
\bea
U\, \ket{0}\ket{0}\ket{X} & = & a\ket{00}\ket{A}+b(\ket{01}
                      +\ket{10})\ket{B}+ c\ket{11}\ket{C}\ \ ,
                  \nonumber \\
                 U\, \ket{1}\ket{0}\ket{X} & = & \tilde a\ket{11}
                          \ket{\tilde A}+\tilde b(\ket{10}
                      +\ket{01})\ket{\tilde B}
                      + \tilde c\ket{00}\ket{\tilde C} \ \ .
                   \label{eq:10}
\eea
For convenience we include all phases in equation
(\ref{eq:10}) into the ancilla states, so that the 
coefficients $a, b, c, \tilde a,\tilde b $ and $\tilde c$
are real and positive. Furthermore the transformation should
not change under renaming the basis, i.e. exchange of $\ket{0}$ 
and $\ket{1}$ -- therefore we have
$a =\tilde a$, $b =\tilde b$ and $c=\tilde c$.
\par
The normalization and  unitarity conditions for equation 
(\ref{eq:10}) read 
\bea
a^2+2b^2+c^2 &=& 1\ \ ,\nonumber \\
a  c \bra{\tilde C}A\rangle + 2b^2 \bra{\tilde B}B\rangle +a c
\bra{\tilde A}C\rangle & = & 0\ \ .
\label{unitary}
\eea
\par
Now we have to determine the free parameters in this 
transformation (coefficients and scalar
products of ancillas) such that the fidelity 
$F=\bra{\psi}\rho^{out}\ket{\psi}$,
where $\ket{\psi}$ is one of the four BB84 states,
is constant and optimal. Here $\rho^{out}$ is the reduced 
density matrix of the first or second bit at the output of the cloner.
\par
It is straightforward to calculate the fidelities
corresponding to the reduced
output density matrices for the four BB84 states.  From their 
equality we find the following constraints:
\bea
F&=&a^2+b^2 \ \ ,\\
F&=&\half (1+a b \, \text{Re}[\bra{\tilde A}B\rangle 
+\bra{\tilde B}A\rangle]
+ b c \,\text{Re}[\bra{\tilde B}C\rangle +\bra{\tilde C}B
\rangle])\ \ ,
\label{fidel}\\
0&=&a b \,\text{Re}[\bra{\tilde A}\tilde B\rangle +\bra{B}A
\rangle ]
+ b c \,\text{Re}[\bra{\tilde B}\tilde C\rangle + \bra{C}B
\rangle ]\ \ .
\label{zero}
\eea
As the scalar products of ancillas are independent parameters
the real part of
which varies  between -1 and +1, we
can maximise the fidelity in equation (\ref{fidel}) to
\be
F= \half (1+ 2 b (a +c)) \ \
\label{max}
\ee
by an appropriate choice of ancillas. Similarly, we can always
fulfill equation (\ref{zero}) by the right choice of ancillas.
So, our task reduces to finding the maximum of the function
\be
F=\half (1+a^2-c^2)\ \ ,
\ee
with the constraint
\be
F=\half+\sqrt{\half(1-a^2-c^2)}(a+c) \ \ .
\ee
This can be done analytically with the help of Lagrange 
multipliers.
The solution for the optimum is
\bea
a&=&\half+\seight \ \ ,\nonumber \\
b&=&\seight \ \ ,\nonumber \\
c&=&\half-\seight  \ \ .
\eea
This solution corresponds to an optimal fidelity of
\be
F^{opt}(1,2)=\half+\seight=0.854 \ \ ,
\ee
which reaches the bound $\tilde F_{pcc}(1,2)=\frac{1}{2}(\tilde{\eta}_{pcc}(1,2)+1)$,
given by equation (\ref{limsup}).
\par
The optimal cloning transformation for the BB84 states
can be written explicitly as follows (we see
that a two-dimensional ancilla is sufficient):
\bea
U\, \ket{0}\ket{0}\ket{X} & = &
  (\half+\seight)\ket{00}\ket{0}+\seight(\ket{01}+\ket{10})
\ket{1}+(\half-\seight)\ket{11}\ket{0} \ \ ,  \nonumber \\
U\, \ket{1}\ket{0}\ket{X} & = & (\half+\seight)\ket{11}
\ket{1}+\seight(\ket{10}+\ket{01})\ket{0}
+ (\half-\seight)\ket{00}\ket{1}.
\label{trafo}
\eea
\par
We still have to show that this transformation leads to the 
same fidelity for {\em any} pure input state taken from the 
equator.
In fact, any unitary transformation of the kind
\bea
U\, \ket{0}\ket{0}\ket{X} & = &
a \ket{00}\ket{0}+b(\ket{01}      +\ket{10})\ket{1}+
c\ket{11}\ket{0} \ \ , \nonumber \\
U\, \ket{1}\ket{0}\ket{X} & = & a\ket{11}
     \ket{1}+b(\ket{10}+\ket{01})\ket{0}
                      + c\ket{00}\ket{1},
\label{trafo1}
\eea
that leads to the same fidelity for the BB84 states
has this property. This can be seen by calculating 
the fidelity when applying the transformation (\ref{trafo1})
to the state given in equation (\ref{state}).
We find
\be
F(\alpha) = (\alpha^4 +\beta^4) a^2+b^2+\alpha^2\beta^2\cdot 
2 c^2 +4 \alpha^2\beta^2 b(a+c) \ \ ,
\ee
which at first glance does not look like a constant, but can 
be shown easily to be independent of $\alpha$  by 
inserting equation (\ref{max}) and the constraints from unitarity,
given in eq. (\ref{unitary}). Thus
we have shown that apart from the four BB84 states our 
cloner (\ref{trafo}) is optimal for {\em any} state from the equator.
\par
It is worth pointing out that there is a link between optimal cloning of  
equatorial qubits and optimal eavesdropping
in the BB84 scheme, see \cite{eave}: the intersection of the
curve for the mutual information between Alice and Bob and 
the curve for the optimal mutual information between Alice and Eve 
occurs at a disturbance $D=1-F$ which corresponds to 
our optimal equatorial
cloning fidelity: if Eve performs a symmetric attack where she 
gets as much information as Bob she cannot find a  better
strategy than applying the best cloner. We could
have actually proved an upper bound for our cloner from a 
contradiction:
let us assume the cloner could have a higher fidelity than 
the one indicated by the intersection of the information curves. 
Then Eve could
use it to eavesdrop and would have found a better  spying
device than the optimal one. Therefore the best cloner cannot have
a higher fidelity than the best symmetric eavesdropping attack.
In this section we have shown  a constructive proof for
the corresponding optimal cloning transformation.

\section{Conclusions }

In this article we have pointed out a connection between optimal cloning
of equatorial qubits and phase estimation. We exploited this connection to
establish an upper bound for the fidelity of a phase covariant
$N\to M$ cloning transformation acting on equatorial qubits. 
Our results 
for this restricted set of inputs are qualitatively similar
to the ones for universal cloning, in the sense that in both cases
the concatenation property holds. Quantitatively our upper bound is
higher than the one for universal cloning, as expected.  
The bound for phase covariant cloning was shown to be reached for $N=1, M=2$
by constructing the optimal cloning transformation explicitly. 
In this particular case we also found a link between phase covariant cloning
and optimal eavesdropping strategies in the quantum cryptographic scheme
BB84.
Finding the  explicit optimal phase covariant cloning 
transformation for general $N$ and
$M$ remains to be achieved.
\par
This work was supported in part by Deutsche Forschungsgemeinschaft under grant 
SFB 407, by Ministero dell'Universit\`a e della Ricerca Scientifica e 
Tecnologica under the project ``Amplificazione e rivelazione di radiazione 
quantistica'' and by the ESF-QIT programme.

\appendix
\section{Map of the phase covariant cloner}
\label{app}

We use the Kraus decomposition \cite{Kra} of
 a CP-map (the map $R T_{N,M}$ in Eq. (\ref{out}) is CP since it is the 
partial trace  of the CP-map $T_{N,M}$)
\begin{equation}
R(T_{N,M}(\proj{\psi}^{\otimes N}))=
\sum_{k}A_{k}\proj{\psi} A_{k}^{\dagger}\label{CP} \ ,
\end{equation}
where $A_{k}$ are operators on $\mathcal{H}$ depending on $N$ and $M$, 
satisfying the condition
\begin{equation}
\sum_{k}A_{k}^{\dagger}A_{k}=\eins \ .
\label{CP1}
\end{equation}

By introducing the following basis for the $\mathbb{C}$-algebra of the 
operators on $\mathcal{H}$
\begin{eqnarray}
& & \sigma_{0}=\frac{1}{2}(\sigma_x+i\sigma_y)\ ,
\quad\sigma_{1}=\frac{1}{2}(\sigma_x-i\sigma_y)\ , 
\\
& & \sigma_{2}=\frac{1}{2}(1+\sigma_z)\ , \ \ \quad\sigma_{3}=
\frac{1}{2}(1-\sigma_z)\ ,
\label{sigmas}
\end{eqnarray}
we can write in general
\begin{equation}
A_{k}=\sum_{\alpha=0}^{3}c_{k}^{\alpha}\sigma_{\alpha}\ ,
\label{CP3}
\end{equation}
with  $c_{k}^{\alpha}\in \mathbb{C}$. It follows that
\begin{eqnarray}
R(T_{N,M}(\proj{\psi}^{\otimes N})) 
& = & \sum_{k}\sum_{\alpha,\beta=0}^{3}c_{k}^{\alpha}c_{k}^{\beta *} 
\sigma_{\alpha} \proj{\psi} \sigma_{\beta}^{\dagger}=\nonumber\\
& = & \sum_{\alpha,\beta=0}^{3}\Gamma^{\alpha\,\beta}\Sigma_{\alpha\,\beta}
(\proj{\psi})\ ,
\label{CP4}
\end{eqnarray}
with $\Sigma_{\alpha\,\beta}(\proj{\psi})\equiv \sigma_{\alpha}\proj{\psi}
\sigma_{\beta}^{\dagger}$ and $\Gamma^{\alpha\,\beta}\equiv
\sum_{k}c_{k}^{\alpha}c_{k}^{\beta *}$.
\par
Imposing
the phase-covariance condition (\ref{covariance}) to the above CP-map
and using (\ref{CP4}) we find
\begin{equation}
\sum_{\alpha,\beta}\Gamma^{\alpha\beta}\Sigma_{\alpha\beta}
(U_{\chi}\proj{\psi} 
U_{\chi}^{*})=\sum_{\alpha,\beta}\Gamma^{\alpha\beta} U_{\chi}
\Sigma_{\alpha\beta}(\proj{\psi}) U_{\chi}^{*}\label{Cov4}\ .
\end{equation}
Writing down explicitly each term of Eq. (\ref{Cov4}) and imposing that the 
equality holds $\forall \chi\in [0,2\pi)$ 
we obtain the following constraints 
on the coefficients $\Gamma^{\alpha\beta}$:
\begin{eqnarray}
& & \Gamma^{01}=\Gamma^{02}=\Gamma^{03}=0\ ,\nonumber\\
& & \Gamma^{10}=\Gamma^{12}=\Gamma^{13}=0\ ,\nonumber\\
& & \Gamma^{20}=\Gamma^{21}=0\ ,\nonumber\\
& & \Gamma^{30}=\Gamma^{31}=0\ .
\label{gamma}
\end{eqnarray}
In order to obtain Eq.(\ref{gamma}) we have written a general density 
matrix in $\mathcal{H}$ as
\begin{equation}
\varrho =\pmatrix{\delta & \gamma\cr \gamma^{*} & 1-\delta\cr}
\label{genericarho}
\ee
with $\delta\in [0,1]$ and $\gamma\in \mathbb{C}$.
The condition (\ref{CP1}) takes the form
\be
\sum_{\alpha,\beta=0}^{3}\Gamma^{\beta\alpha}\sigma_{\alpha}^{\dagger}
\sigma_{\beta}=\eins \ ,
\label{CPII1}
\ee
which gives
\begin{equation}
\Gamma^{11}=1-\Gamma^{22},\quad\quad\Gamma^{00}=1-\Gamma^{33}\ .
\label{CPII7}
\end{equation}
Note that $\Gamma^{\alpha\alpha}=\sum_{k}|c_{k}^{\alpha}|^{2}\geq 0\quad
\forall \alpha$ and $\Gamma^{\alpha\beta}=(\Gamma^{\beta\alpha})^{*}$.
Using (\ref{CPII7}) we have $0\leq\Gamma^{\alpha\alpha}\leq 1$ and 
$|c_k^\alpha|\leq 1$ $\forall
\alpha$, from which we obtain
\begin{equation}
|\Gamma^{32}|^{2}=|\sum_{k}c_{k}^{3}c_{k}^{2*}|^{2}
\leq
\sum_{k}|c_{k}^{3}
c_{k}^{2*}|^{2}\leq\Gamma^{22}\Gamma^{33}\leq 1 \ .
\end{equation}
Using the conditions (\ref{gamma}) and (\ref{CPII7}) we can now write Eq. 
(\ref{CP4}) in matrix form as follows 
\begin{equation}
R(T_{N,M}(\proj{\psi}^{\otimes N}))=
\pmatrix{(1-\Gamma^{33})(1-\delta)+\Gamma^{22}\delta & \gamma\;
\Gamma^{32}\cr \gamma^{*}(\Gamma^{32})^* & 
(1-\Gamma^{22})\delta+\Gamma^{33}(1-\delta)\cr}\label{RT3}
\end{equation}
Let us now use the notation $\eta_{xy}\equiv|\Gamma^{32}|$, $\varphi\equiv
\arg{(\Gamma^{32})}$ and 
$\eta_z=(\Gamma^{33}+\Gamma^{22}-1)$.
Note that $0\leq\eta_{xy}\leq 1$, $-1\leq \eta_z\leq 1$,
 and $\eta_{xy,z}=\eta_{xy,z}(N,M)$: the dependence on $N$ and $M$
 is included in the  coefficients $c_{k}^{\alpha}$.
 \par 
Comparing the Bloch vector of an input generic qubit $\vec s^{in}=(2|\gamma|
\cos{\phi},-2|\gamma|\sin{\phi},2\delta-1)$ 
where $\phi=\arg{(\gamma)}$ with the 
Bloch vector of the one-particle
 reduced density matrix of the output $\vec s^{out}=
(2\eta_{pcc}|\gamma|\cos{(\phi+\varphi)},-2\eta_{pcc}
|\gamma|\sin{(\phi+\varphi)},
s_{z}^{in}\eta_z+(\Gamma^{22}-\Gamma^{33}))$, we notice 
that for 
\be
\varphi=0\quad\quad\quad \text{and} \ \ 
\Gamma^{22}=\Gamma^{33}\label{condizioni}
\ee
the map $T_{N,M}$ 
is completely determined by the factors  $\eta_{xy}(N,M)$ and 
$\eta_z(N,M)$: $\eta_{xy}(N,M)$ 
describes the shrinking of the Bloch vector in the $xy$ plane, 
while $\eta_z$ gives 
the shrinking along the $z$-axis.
For initial equatorial qubits ($\delta=1/2$, $\gamma=e^{i\phi}/2$)
we find with the conditions (\ref{condizioni}): 
\begin{eqnarray}
R(T(\proj{\psi_\phi}^{\otimes N})) 
& = & \frac{1}{2}\pmatrix{1 & \eta_{xy}(N,M)e^{i\phi}\cr 
\eta_{xy}(N,M)e^{-i\phi} & 1\cr}\\\nonumber\\
& = & \eta_{xy}(N,M)\proj{\psi_\phi}+\frac{1}{2}[1-\eta_{xy} (N,M)] \eins
\label{RT10} 
\end{eqnarray}
i.e. the action of of the phase-covariant cloner $T_{N,M}$
on equatorial qubits 
is completely determined by the shrinking factor $\eta_{xy}(N,M)$ 
in the $xy$ plane.

Let us now show that without loss of generality we can impose the 
conditions (\ref{condizioni}) to describe the map of an optimal 
phase covariant
cloner for equatorial qubits.
For $\varphi\ne 0$ the fidelity for equatorial qubits is given by
\begin{eqnarray}
F_{pcc}(N,M)& = & |\langle\psi_{\phi}|R(T_{N,M}(\proj{\psi_\phi}^{\otimes N}))
|\psi_{\phi}\rangle|^{2}=
\nonumber\\ & = & \frac{1}{2}(1+\eta_{xy}(N,M)\cos{\varphi})\label{phi1}
\end{eqnarray}
By definition the optimal cloner ${T}_{N,M}$ is the one which maximizes 
$F_{pcc}(N,M)$. From (\ref{phi1}) we see that maximizing $F_{pcc}(N,M)$ 
is equivalent to setting 
$\varphi=0$ and maximizing $\eta_{xy}(N,M)$, which is independent of
$\varphi$.

Let us now analyse the condition $\Gamma^{22}=\Gamma^{33}$. 
Let us suppose that we can find an optimal 
phase-covariant cloner ${T}_{N,M}$ with $\eta_{xy}^{opt}(N,M)$ and 
$\Gamma^{22}\ne\Gamma^{33}$.
From the explicit form of
 $\sigma_{\scriptsize{2}}$ and $\sigma_{\scriptsize{3}}$,
 given in (\ref{sigmas}), one can see that  
  renaming the basis (i.e. exchanging $\ket{0}\leftrightarrow \ket{1}$) is 
equivalent to exchanging 
$\sigma_{\scriptsize{2}}\leftrightarrow\sigma_{\scriptsize{3}}$
and $\sigma_{\scriptsize{0}}\leftrightarrow\sigma_{\scriptsize{1}}$,
 while leaving the basis vectors unchanged.
The exchange  
$2\leftrightarrow 3$
leaves $\eta_{xy}(N,M)$ and  $\eta_{z}(N,M)$ invariant. 
Now consider a cloner ${\hat T}_{N,M}$ such that its single particle 
reduced density  operator is 
the matrix $R({T}_{N,M}(\proj{\psi_\phi}^{\otimes N}))$ written in the form 
(\ref{CP4}) with the exchange $2\leftrightarrow 3$. The map ${\hat T}_{N,M}$ 
must also be 
optimal: in fact optimality of ${T}_{N,M}$ cannot depend on the particular 
choice of the basis, and the fidelity (\ref{phi1}) is invariant under the
 exchange $2\leftrightarrow 3$.
Now consider the cloner described by the map $T_s=\frac{1}{2}
({T}_{N,M}+{\hat T}_{N,M})$. 
This cloner has
the same shrinking factor $\eta_{xy}^{opt}(N,M)$ for equatorial qubits.
Therefore we can always construct an optimal cloner with 
$\Gamma^{22}=\Gamma^{33}$.

\end{document}